\documentclass[a4paper]{jpconf}

\newcommand{\be}{\begin{equation}}
\newcommand{\ee}{\end{equation}}

\newcommand{\bea}{\begin{eqnarray}}
\newcommand{\eea}{\end{eqnarray}}
\newcommand{\dis}{\displaystyle}

\usepackage{graphicx}
\begin{document}
\title{Analysis of Realized Volatility for Nikkei Stock Average on the Tokyo Stock Exchange}

\author{Tetsuya Takaishi${}^1$ and Toshiaki Watanabe${}^2$}

\address{$~^1$Hiroshima University of Economics, Hiroshima 731-0192, Japan \\
$~^2$Institute of Economic Research, Hitotsubashi University, Tokyo 186-8603, Japan
}

\ead{tt-taka@hue.ac.jp}

\begin{abstract}
We calculate realized volatility of the Nikkei Stock Average (Nikkei225) Index on the Tokyo Stock Exchange
and investigate the return dynamics.
To avoid the bias on the realized volatility from the non-trading hours issue 
we calculate realized volatility separately in the two trading sessions, i.e. morning and afternoon, of the Tokyo Stock 
Exchange and find that the microstructure noise decreases the realized volatility at small sampling frequency.
Using realized volatility as a proxy of the integrated volatility we standardize returns in the morning and afternoon sessions and 
investigate the normality of the standardized returns by calculating variance, kurtosis and 6th moment.
We find that variance, kurtosis and 6th moment are consistent with those of the standard normal distribution,
which indicates that the return dynamics of the Nikkei Stock Average are well described by 
a Gaussian random process with time-varying volatility.

\end{abstract}

%------------------------------------------------------------------------
% begin document
%------------------------------------------------------------------------

\section{Introduction}
Statistical properties of asset  returns have been extensively studied and 
it is found that asset price returns show some universal properties that are 
not explained well in the framework of the standard Brownian motion. 
The universal properties are now classified as the stylized facts of asset price returns
which includes: fat-tailed return distributions, volatility clustering, long autocorrelation time in
absolute returns and so on\cite{Cont}.
To explain the fat-tailed return distribution Mandelbrot introduced 
a class of stable processes such as stable Paretian\cite{Mandelbrot}.
An alternative idea to explain the asset price dynamics was given by Clark who 
related the volatility variation to volume and suggested to use 
the subordinated process to the asset price dynamics\cite{Clark}.
The idea for the asset price dynamics by Clark is also called
the mixture of distributions hypothesis (MDH).
The return process with the MDH does not conflict with major properties observed in asset returns, e.g.
volatility clustering, fat-tailed return distributions.
Under the MDH, the asset return at discrete time $t$ can be described by
$r_t=\sigma_t \epsilon_t$, where $\sigma_t^2$ is a variance of the Gaussian distribution and $\epsilon_t$ is 
a standard normal random variable, and this indicates that the asset return process is viewed as  
a Gaussian random process with time-varying variance (volatility). 
Let $p(r|\sigma_t^2)$ be the conditional return distribution with $\sigma_t^2$
and $p(\sigma_t^2)$ the probability distribution of volatility.  
The unconditional return distribution $p(r)$ from this process is obtained by  integrating 
the conditional return distribution and the probability distribution of volatility 
with respect to volatility, i.e. $ p(r)=\dis \int  p(r|\sigma_t^2) p(\sigma_t^2) d\sigma_t^2$. 
Empirical studies suggested that
the volatility distributions might be described by the inverse gamma distribution or log-normal
distribution\cite{Clark,Praetz,Beck,TakaishiDist} with which the unconditional return distributions become fat-tailed distributions.
 
The verification of the MDH can be made by testing the returns standardized by 
volatility, $r_t/\sigma_t$.
Under the MDH, the standardized returns should behave  as $r_t/\sigma_t\sim \epsilon_t$, i.e. standard normal random variables.
This test has been conducted in the literature\cite{SR1,SR2,SR3,SR4,SR5,Fleming,TakaishiSR1,TakaishiSR2} 
and it is shown that the MDH is hold for many cases.
For this test it is crucial to use a precise measure of volatility. 
So far the most precise measure is the realized volatility\cite{RV} constructed from high-frequency price data.
Under ideal circumstances the realized volatility goes to the integrated volatility in the limit of the infinite sampling frequency.
However such circumstances are usually violated and the realized volatilities in empirical cases are biased. 
There exist two main sources of bias: microstructure noise and non-trading hours.   
Thus to verify the standard normality of the returns standardized by realized volatility it is important to 
control such biases.
In \cite{TakaishiSR1}, in order to avoid bias from non-trading hours the MDH is tested separately in morning and afternoon sessions
for individual Japanese stocks 
and it is shown that after removing the finite-sample effect\cite{Peter} the return dynamics becomes consistent with the MDH\cite{TakaishiSR2}.
In this paper we focus on the realized volatility of the Nikkei Stock Average index and investigate whether the MDH can also apply 
for the price dynamics of the Nikkei Stock Average index.

\section{Realized Volatility}
We assume that the logarithmic price process $\ln p(s)$ follows
a continuous time stochastic diffusion,
\be
d\ln p(s) =\tilde{\sigma}(s)dW(s),
\label{eq:SD}
\ee
where $W(s)$ stands for a standard Brownian motion  and $\tilde{\sigma}(s)$ is a spot volatility at time $s$.
The assumption of a Brownian motion for the logarithmic price process is often used to model stock prices  as in 
the Black-Scholes model\cite{Black,Merton}  and is most widely used to investigate stock price process. 
Using $\tilde{\sigma}(s)$, 
the integrated volatility from $t$ to $t+h$ is defined by 
\be
IV_h(t)= \sigma_h^2(t) =\int_{t}^{t+h}\tilde{\sigma}(s)^2ds,
\label{eq:int}
\ee
where $h$ stands for the interval to be integrated, e.g. $h= 1$ corresponds to one day for the daily integrated volatility.

The realized volatility is designed as a model-free estimate of volatility and constructed as a sum
of squared returns\cite{RV}.
The realized volatility $RV_t$ at time $t$ is given by 
\be
RV_t =\sum_{i=1}^{n} r_{t+i\Delta}^2,
\label{eq:RV}
\ee
where $n$ is the number of returns at sampling frequency $\Delta$, given by $n=h/\Delta$.
The returns sampled at $\Delta$ are given by log-price difference,
\be
r_{t+i\Delta}=\ln P_{t+i\Delta}-\ln P_{t+(i-1)\Delta},
\label{eq:Ret}
\ee
for $i=1,\dots,n$.
Eq.(\ref{eq:RV}) goes to the integrated volatility defined by eq.(\ref{eq:int}) in the limit of $n \rightarrow \infty$.
However what we observe in the real financial markets is not the returns given by eq.(\ref{eq:Ret}).
The asset prices observed in the real financial markets are contaminated with the microstructure noise originating from discrete trading,
bid-ask spread and so on.
Following Zhou\cite{Zhou} let us assume that the microstructure noise introduces independent noises
and 
the log-price observed in financial markets is given by 
\be
\ln P^{*}_t =  \ln P_t +\xi_t,
\label{eq:independent}
\ee
where $\ln P_t^{*}$ is the observed log-price in the markets which consists of the true log-price $\ln P_t$ and
noise $\xi_t$ $\sim N(0,\omega^2)$.
Under this assumption the observed
return $r^{*}_t$ is given by
\be
r^{*}_t=r_t +\eta_t,
\ee
where $\eta_t=\xi_{t}-\xi_{t -\Delta}$.
The realized volatility  $RV^{*}_t$ actually observed at the financial market is obtained
as a sum of the squared returns $r^{*}_t$,
\bea
RV_t^{*}& = &\sum_{i=1}^{n} (r^{*}_{t+i\Delta})^2,  \\
   & =& RV_t + 2\sum_{i=1}^n r_{t+i\Delta}\eta_{t+i\Delta} + \sum_{i=1}^n \eta_{t+i\Delta}^2.
\label{eq:rvnoise}
\eea
After averaging  $RV^{*}_t$, we find that the bias in  $RV^{*}_t$ appears as $\sum_{i=1}^n \eta_{t+i\Delta}^2$
which corresponds to $\sim 2n\omega^2$.
Thus with independent noise  
the $RV^{*}_t$ diverges as $n \rightarrow \infty$.
Another bias appeared in the realized volatility is due to the existence of 
the non-trading hours on the real financial markets.
At the Tokyo Stock Exchange  domestic stocks are traded in
the two trading sessions: (1) morning trading session (MS) from 9:00 to 11:00.
(2) afternoon trading session (AS) from 12:30 to 15:00.
If we calculate the daily realized volatility without including returns during
the non-traded periods it can be underestimated.

In order to avoid the non-trading hours issue
we consider two realized volatilities:(i) $RV_{MS}$, realized volatility in the morning session and
(ii) $RV_{AS}$, realized volatility in the afternoon session.
Since these realized volatilities are calculated separately
each volatility does not have bias due to the non-trading hours issue.

\section{Normality of the Standardized Return}

In this study we analyze the high-frequency data of Nikkei Stock Average (Nikkei225) index 
from May 1, 2006 to December 30, 2009, that corresponds to 900 working days. 
Figure 1 shows return time series of the Nikkei Stock Average in the different 
time zones of the Tokyo Stock Exchange.

\begin{figure}
\vspace{10mm}
\centering
\includegraphics[height=9cm]{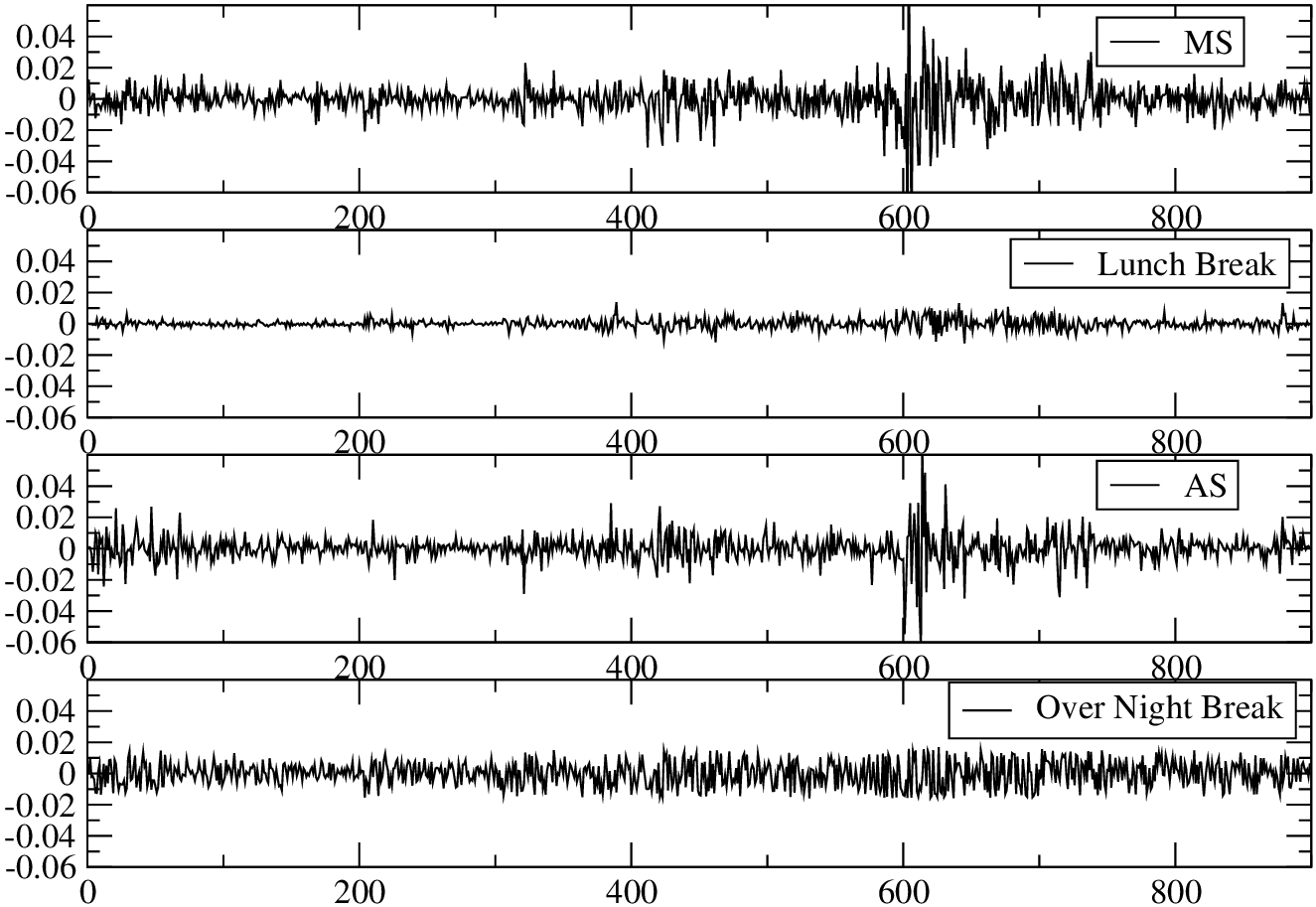}
\caption{
Return time series in different time zones from May 1, 2006 to Dec. 30, 2009.
}
%\vspace{-6mm}
%\label{fig:corr}
\end{figure}

In each zone the return is calculated by the log-price difference between
the beginning and end prices in the corresponding zone.
For instance let $R_{MS,t}$ be the return in the morning session at day $t$.
$R_{MS,t}$ is given by $\ln P^o_{MS,t}-\ln P^c_{MS,t}$, where $P^o_{MS,t}$ is 
the opening price of the morning session and $P^c_{MS,t}$ the closing price of the morning session. 
Similarly  the return in the afternoon session  $R_{AS,t}$ is given by $\ln P^o_{AS,t}-\ln P^c_{AS,t}$.
Moreover the return in the lunch break (LB),  $R_{LB,t}$ is given by $\ln P^c_{MS,t}-\ln P^o_{AS,t}$ 
and  the return in the overnight break (ON),  $R_{ON,t}$ is given by $\ln P^o_{MS,t}-\ln P^c_{AS,t-1}$.
As seen in the figure returns in the lunch break and overnight break also vary.  
Variation of returns in the lunch break is small. This is because the lunch break duration takes  only 90min.
On the other hand considerable variation of returns is seen in the overnight break which has 18 hrs duration.
Since returns in both breaks can vary
the (daily) realized volatility without including the data in both breaks can not be accurate.

\begin{figure}
\vspace{5mm}
\centering
\includegraphics[height=9cm]{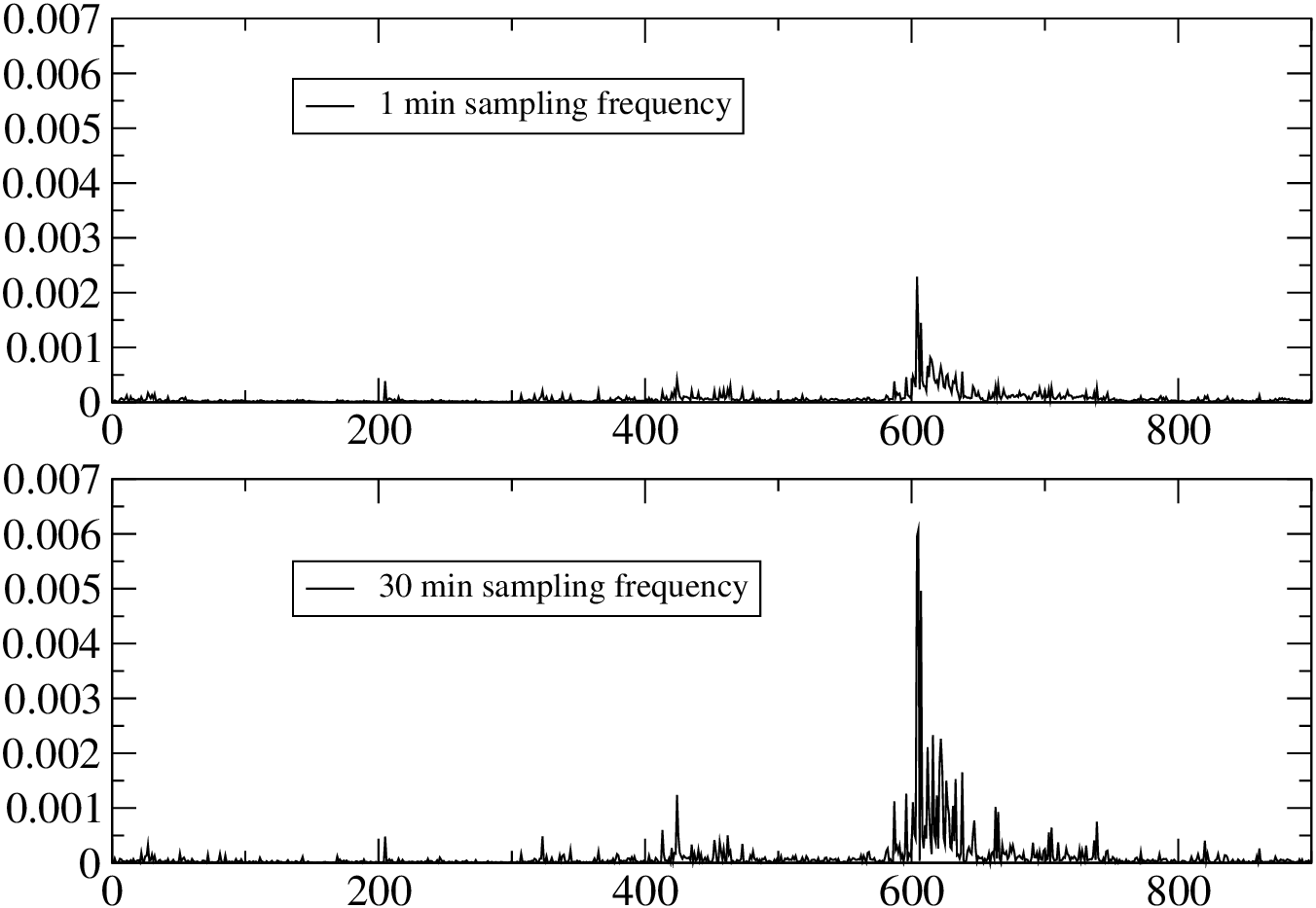}
\caption{
Time series of the realized volatility in the MS from May 1, 2006 to Dec. 30, 2009.
}
%\vspace{-6mm}
%\label{fig:corr}
\end{figure}

\begin{figure}
\vspace{10mm}
\centering
\includegraphics[height=9cm]{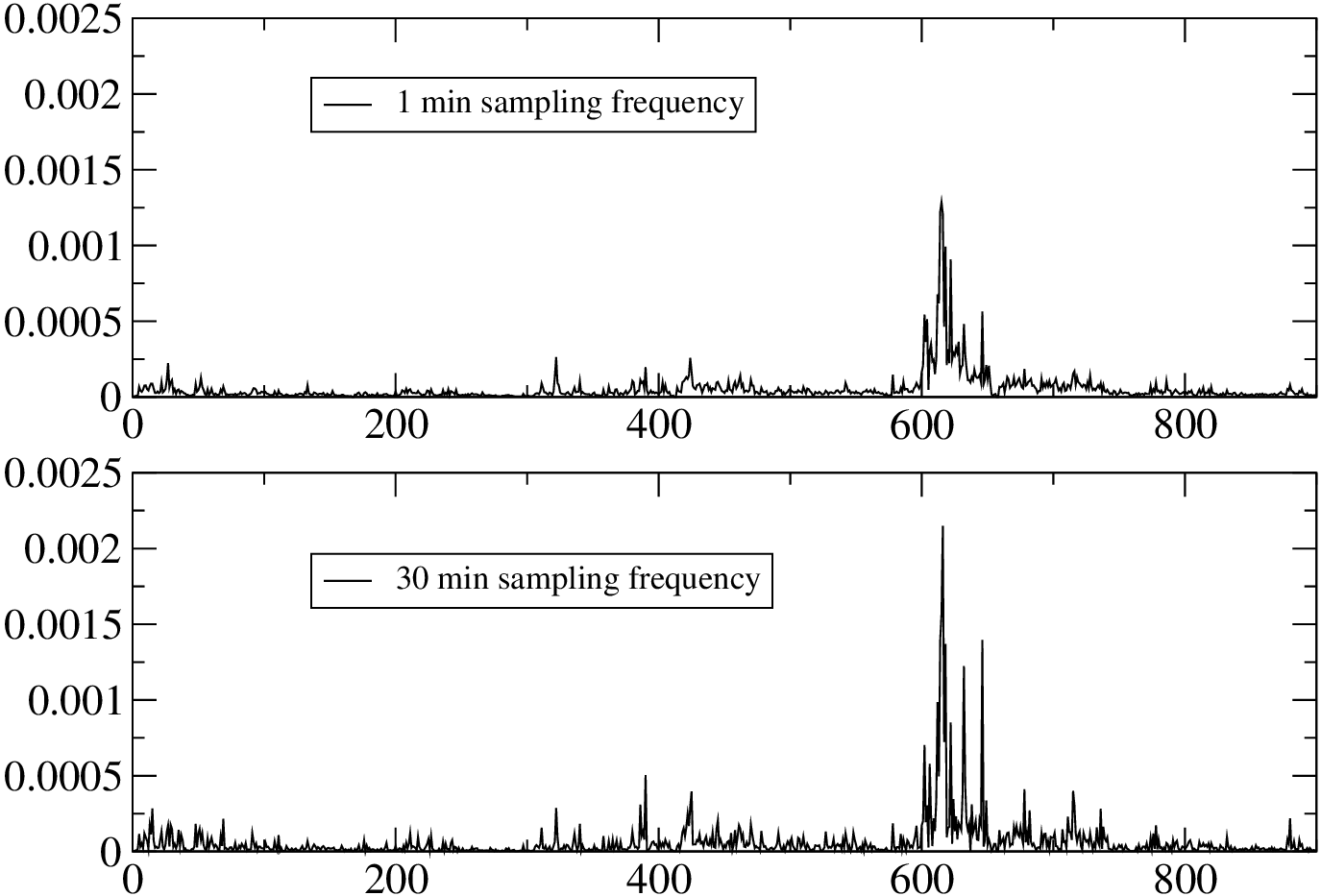}
\caption{
Time series of the realized volatility in the AS from May 1, 2006 to Dec. 30, 2009.
}
%\label{fig:corr}
\end{figure}

In order to avoid the non-trading hours issue
we calculate two realized volatilities in the morning and afternoon session separately.
Let $RV_{MS,t}(RV_{AS,t})$ be the realized volatility in the morning ( afternoon ) session.
Then we analyze standardized returns separately in each session.
$RV_{MS,t}$ is given by 
\be
RV_{MS,t}=\sum^n_{i=1} r^2_{MS,t,\Delta,i},
\ee
where $r^2_{MS,t,\Delta,i}$ is the i-th intraday return in the MS on day $t$, sampled at $\Delta$ min.
Similarly, $RV_{AS,t}$ is given by
\be
RV_{AS,t}=\sum^n_{i=1} r^2_{AS,t,\Delta,i}.
\ee
Here $n$ is the number of returns generated at sampling frequency $\Delta$ during trading sessions.
We calculate the realized volatility for $\Delta=(1,2,...,40)$.
Figure 2 and 3 show the realized volatility in each trading session at $\Delta=1$ and 30 as representative ones.

In figure 4 we show the average of realized volatility as a function of the sampling frequency $\Delta$.
Such plot is also called volatility signature plot\cite{VSP} 
and visualizes the effect of the microstructure noise.
It seems that the  average realized volatility decreases with the sampling frequency $\Delta$.
This is the opposite result to the expectation of eq.(\ref{eq:rvnoise}) and empirical results of individual Japanese stocks\cite{TakaishiSR1} 
where the realized volatility diverges as the sampling frequency $\Delta$ decreases.
Thus for the Nikkei Stock Average the assumption of the independent noise of eq.(\ref{eq:independent}) does not apply.

\begin{figure}
\vspace{10mm}
\centering
\includegraphics[height=9cm]{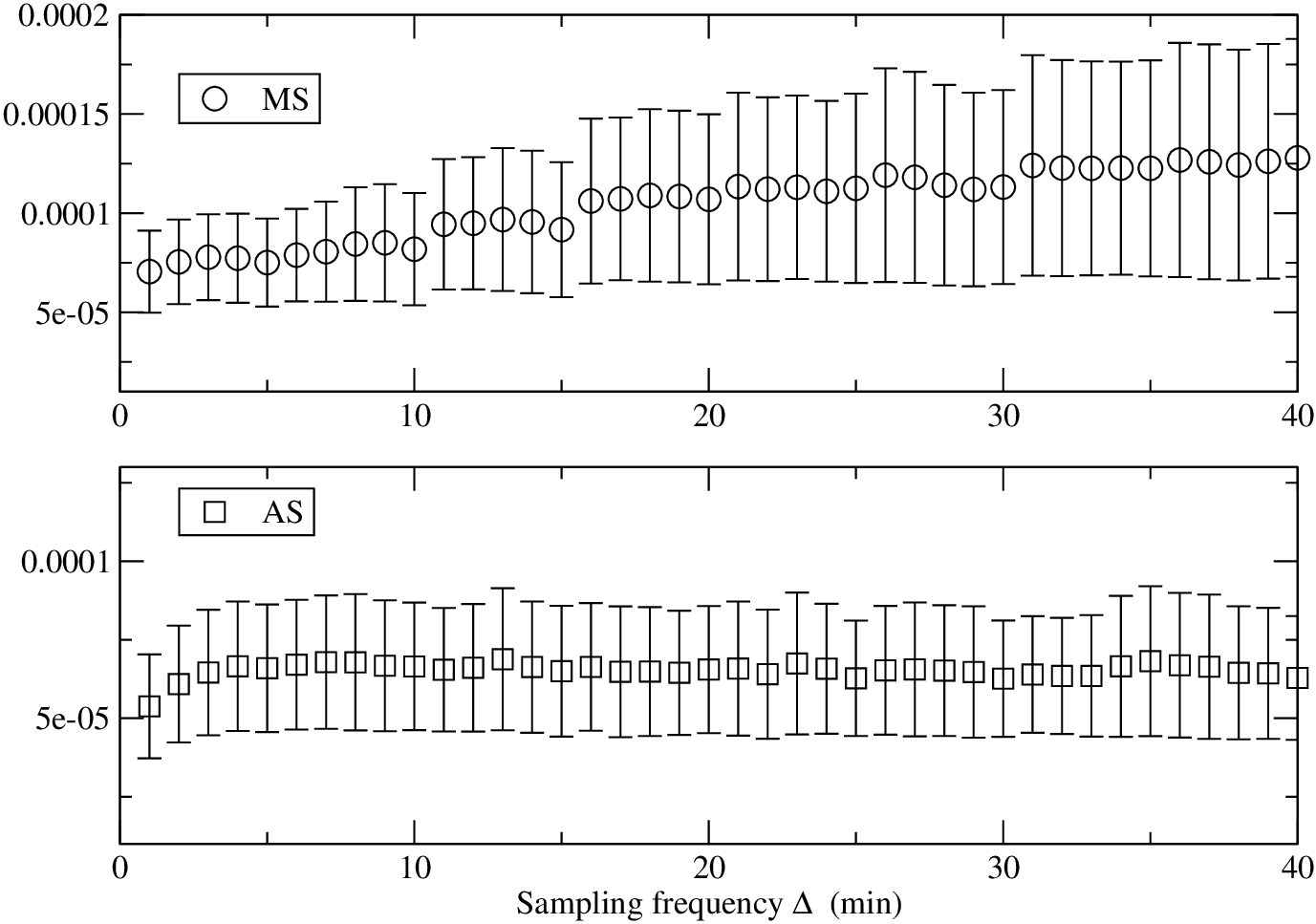}
\caption{
Average of the realized volatility as a function of sampling frequency $\Delta$.
}
%\vspace{-6mm}
%\label{fig:corr}
\end{figure}

\section{Normality of Standardized Returns}

According to the MDH, let us assume that $R_{MS,t}$ and $R_{MS,t}$ are   described by
\be 
R_{MS,t}=\sigma_{MS,t} \epsilon_t,
\label{eq:L1}
\ee
\be
R_{AS,t}=\sigma_{AS,t} \epsilon^{\prime}_t,
\label{eq:L2}
\ee
respectively.
$\sigma_{MS,t}^2 (\sigma_{AS,t}^2)$ is  
an integrated volatility in the morning (afternoon)
and $\epsilon_t$ and $\epsilon^\prime_t$ are independent standard normal random variables $\sim N(0,1)$.
Substituting the realized volatility for the integrated volatility
we standardize $R_{MS,t}$ and $R_{MS,t}$ as
$R_{MS,t}/RV^{1/2}_{MS,t}$ and $R_{AS,t}/RV^{1/2}_{AS,t}$.
These standardized returns are expected to be standard normal variables provided that
eqs.(\ref{eq:L1})-(\ref{eq:L2}) are hold.
However for the realized volatility constructed from the finite number of intraday returns
the standardized return receives the finite-sample effect.
Let $R_s$ be a standardized return.
The distribution of the standardized return $R_s$ is given by\cite{Peter}
\be
f(R_s)=\frac{\Gamma(n/2)}{\sqrt{\pi n}\Gamma((n-1)/2)}\left (1-\frac{R_s^2}{n}\right )^{(n-3)/2}\times 
I(\sqrt{n}\le R_s \le \sqrt{n}),
\ee
where $n$ is the number of returns used to construct the realized volatility. 
The indicator function $I(X)$ means that $I(X)=1$ if $X$ is true and otherwise $I(X)=0$. 
Under this distribution the even moments of $R_s$ are also calculated to be\cite{Peter}
\be
m^{2k}=\frac{n^k(2k-1)(2k-3)\dots 1}{(n+2k-2)(n+2k-4)\dots n}.
\label{eq:theory}
\ee
Actually this finite-sample effect on the standardize return has been observed in empirical results\cite{SR4,SR5,TakaishiSR2} 
and also in the spin model simulation\cite{Spin}.

Figure 5 shows the variance of the standardized returns as a function of sampling frequency $\Delta$.
Here note that from eq.(\ref{eq:theory}) $m^2$ always takes one, i.e. the variance has no finite-sample effect.
As seen in the figure the variance is very close to one through every sampling frequency.  
Only we see an increase at small $\Delta$, which may indicate the effect of the microstructure noise.

\begin{figure}
%\vspace{10mm}
\centering
\includegraphics[height=9cm]{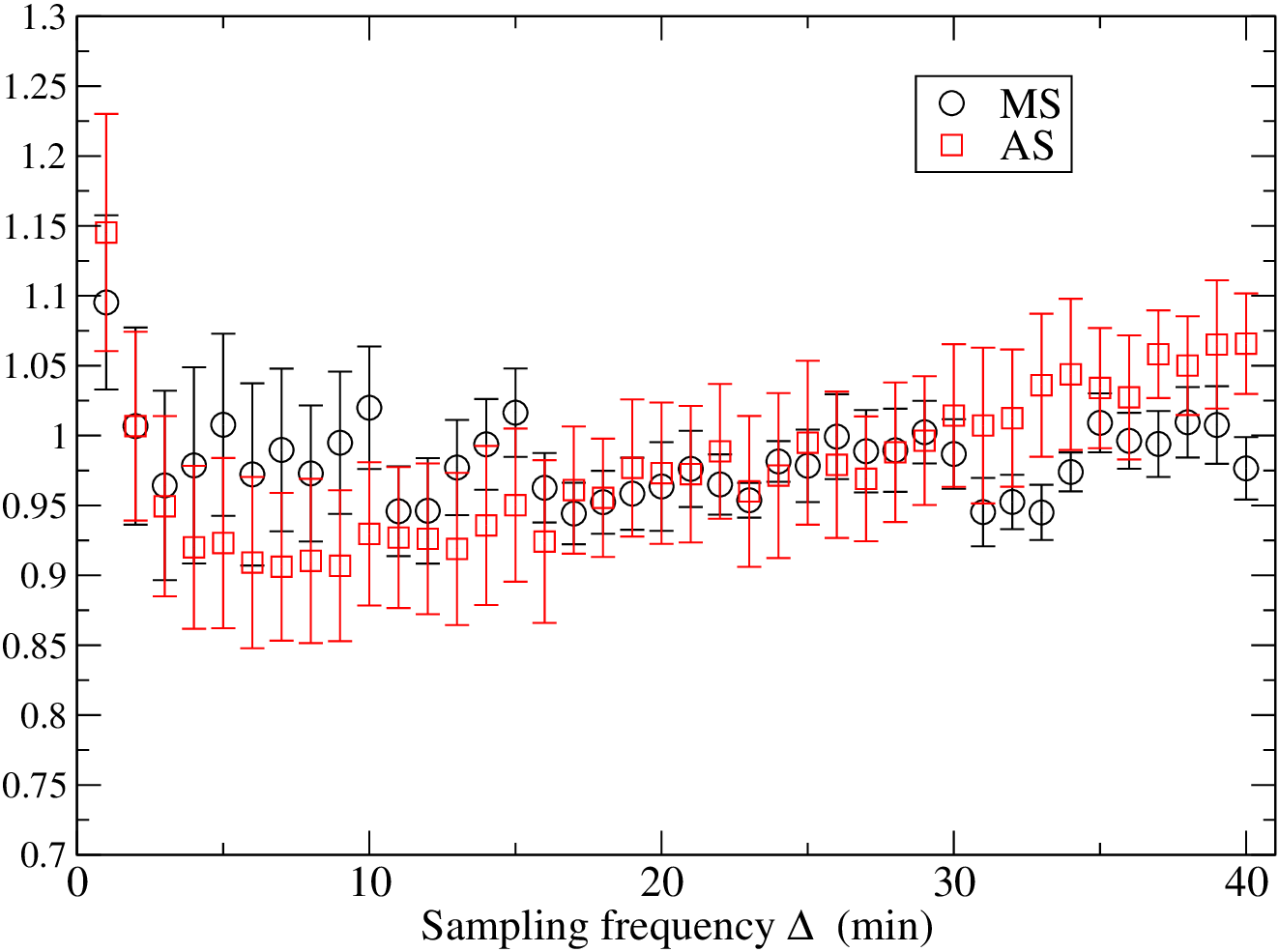}
\caption{
Variance of the standardized returns as a function of sampling frequency $\Delta$.
}
%\vspace{-6mm}
%\label{fig:corr}
\end{figure}

Figure 6 shows the kurtosis as a function of the sampling frequency $\Delta$.
It is seen that the kurtosis decreases as the sampling frequency $\Delta$ increases
as expected from the functional form of the finite-sample effect in eq.(\ref{eq:theory}).
This behavior is also observed for individual stocks on the Tokyo Stock Exchange\cite{TakaishiSR2}. 
The solid lines in the figure are results fitted to a fitting function of
$K(1-\frac2{B_4/\Delta +2})$ where $K$ and $B_4$ are fitting parameters. 
$K$ corresponds to the kurtosis in the limit of $\Delta \rightarrow 0$.
The fitting results are listed in Table 1.
We obtain $K=2.42 (2.86)$ for the MS(AS).
These values are close to 3 expected for the standard normal distributions although the value 
of the MS deviates slightly from 3.

\begin{figure}
\vspace{15mm}
\centering
\includegraphics[height=9cm]{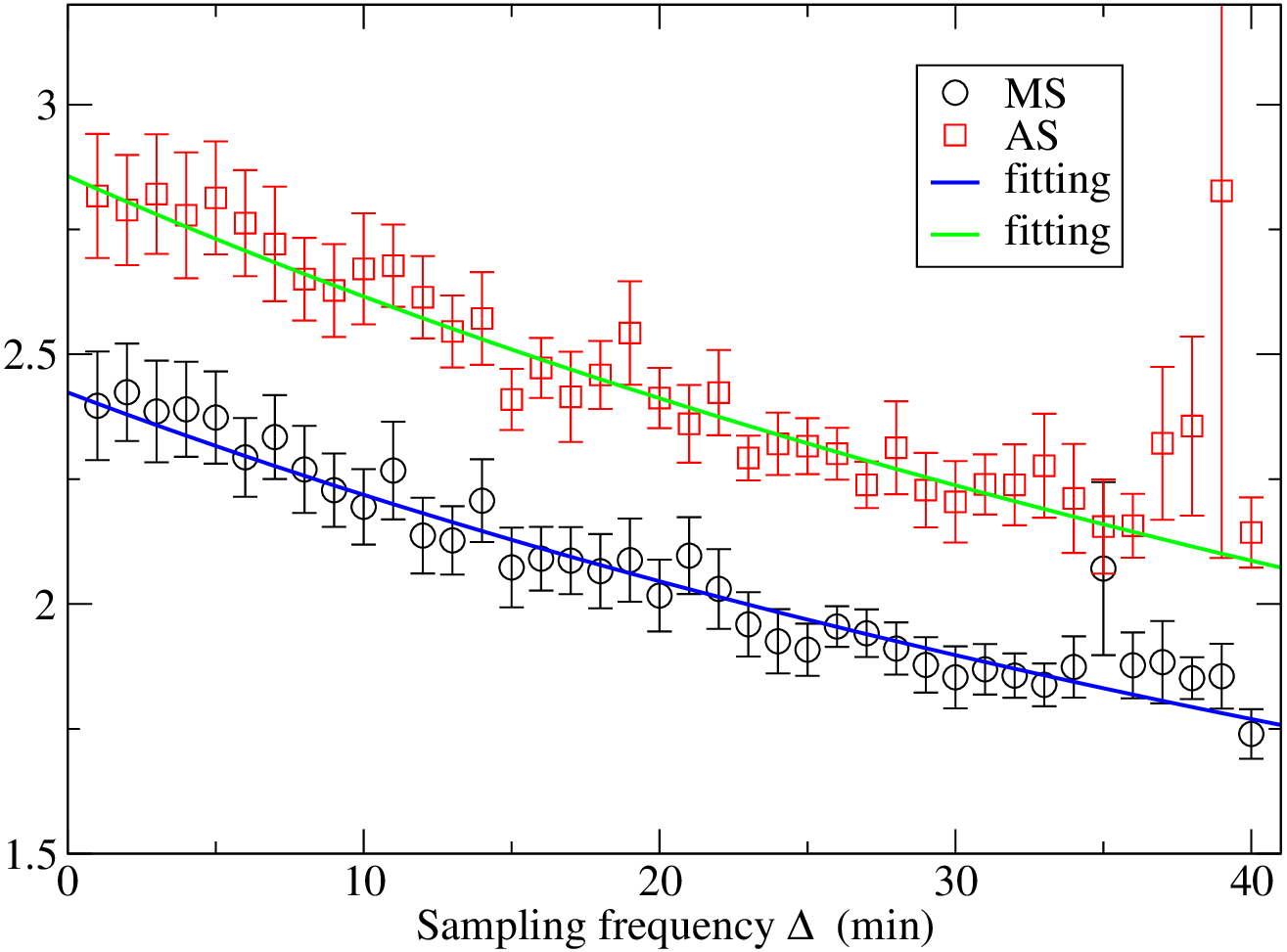}
\caption{
Kurtosis of the standardized returns as a function of sampling frequency $\Delta$.
}
%\vspace{-6mm}
%\label{fig:corr}
\end{figure}

We further analyze the 6th moment of the standardized returns.
The higher moments of the standardized returns have been investigated in the spin financial model
where no microstructure noise exists and it has been shown that
the behavior of the higher moments with varying $\Delta$ is consistent with 
the finite-sample effect described by eq.(\ref{eq:theory})\cite{Spin}.

Figure 7 shows  the 6th moment of the standardized returns as a function of $\Delta$.
The 6th moment increases as the sampling frequency $\Delta$ decreases and 
it seems that the 6th moment approaches the theoretical value, i.e. 15 as $\Delta$ goes to zero. 
We assume that the 6th moment is fitted by a function of $\frac{M_6L^2}{(L+4)(L+2)}$ where $L=B_6/\Delta$,
and $M_6$ and $B_6$ are fitting parameters. 
For this fitting function the value of the 6th moment at $\Delta=0$ is obtained by $M_6$.
The fitting results are listed in Table 1 and we obtain $M_6=9.17(11.6)$ for the MS(AS).

\begin{figure}
\vspace{5mm}
\centering
\includegraphics[height=9cm]{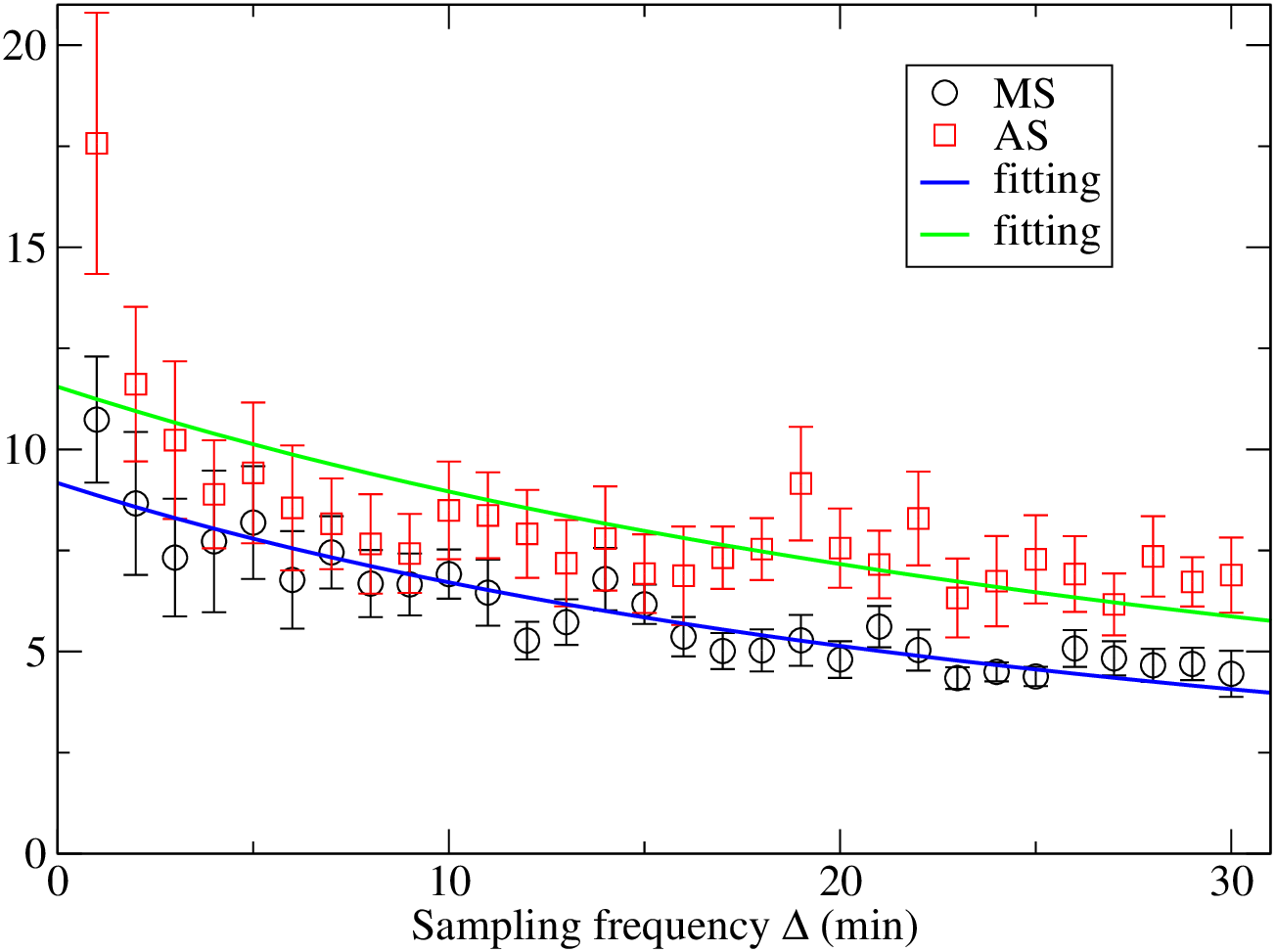}
\caption{
6th moment of the standardized returns as a function of sampling frequency $\Delta$.
}
%\vspace{-6mm}
%\label{fig:corr}
\end{figure}

\begin{table}[t]
%\footnotesize
%\vspace{-2mm}
\centering
\caption{Fitting results. 
The values in parentheses show the theoretical values expected from the standard normal distribution.
The moment $m^{2k}$ of the standard normal distribution is given by $m^{2k}=\int x^{2k}P(x)dx$ where $P(x)$ is the standard normal distribution.
Then we obtain $m^2=1, m^4=3$ and $m^6=15$. Kurtosis is $m^4/(m^2)^2=3$.
It is found that $K$ and $M_6$ extracted at $\Delta=0$ come close to their theoretical values, which indicates that
the results are consistent with the MDH.
}

\begin{tabular}{c|ll|ll}
\hline
       &  Kurtosis (3)  &       &  6th moment (15)             \\
       & $K$         & $B_4$ &  $M_6$       & $B_6$           \\
\hline
MS     &  2.42       & 216.5      & 9.17   &  176.2       \\
AS    &  2.86       & 216.7      & 11.6   &  219.7       \\
\hline
\end{tabular}
%\vspace{-2mm}
\end{table}

\section{Conclusion}
We calculated realized volatility using high-frequency data of the Nikkei Stock Average (Nikkei225) Index on the Tokyo Stock Exchange.
To avoid the bias from the non-trading hours issue
realized volatility was calculated separately in the morning and afternoon trading sessions of the Tokyo Stock
Exchange.
As seen in figure 4 it is found that for the Nikkei Stock Average the microstructure noise decreases the realized volatility at small sampling frequency.
This finding is the opposite result to the microstructure noise appeared on individual Japanese stocks\cite{TakaishiSR1}.
Using realized volatility as a proxy of the true volatility we standardized returns in the morning and afternoon sessions and
investigated the normality of the standardized returns by calculating variance, kurtosis and 6th moment.
It is found that variance, kurtosis and 6th moment are consistent with those of the standard normal distribution,
which indicates that the return dynamics of the Nikkei Stock Average are well described by
Gaussian random process with time-varying volatility, expected from the mixture of distributions hypothesis.

\section*{Acknowledgements}
Numerical calculations in this work were carried out at the
Yukawa Institute Computer Facility and the facilities of the
Institute of Statistical Mathematics. This work was supported
by JSPS KAKENHI Grant Number 25330047.

\end{document}